%% file: hadron2011-bonomi.tex
\documentclass[a4paper,11pt]{article}

\usepackage{contribution}
\usepackage{multirow}

\input{econfmacros}
\input{contributionmacros}

\begin{document}

\input{bonomi}

\end{document}

%% file: econfmacros.tex
\newcommand{\weblink}[2][]{%
    \ifthenelse{\equal{#1}{}}%
    {\textnormal{\url{#2}}}%
    {\textnormal{\href{#2}{#1}}}%
}

\def\beq{\begin{equation}}
\def\eeq#1{\label{#1}\end{equation}}
\def\eeqn{\end{equation}}

\def\beqa{\begin{eqnarray}}
\def\eeqa#1{\label{#1}\end{eqnarray}}
\def\eeqan{\end{eqnarray}}

\let\bar=\overbar

\def\Dslash{\not{\hbox{\kern-4pt $D$}}}
\def\dslash{\not{\hbox{\kern-2pt $\del$}}}

\def\msb{{\bar{\ssstyle M \kern -1pt S}}}

%% file: contributionmacros.tex
\newcommand{\contribution}[7][]{%
  \clearpage
  \thispagestyle{plain}
  \ifthenelse{\equal{#1}{}}
  {\hypersetup{pdftitle={#2}}}
  {\hypersetup{pdftitle={#1}}}
  \hypersetup{pdfauthor={{#3} {#4}}}
  {\centering\normalfont\LARGE\bfseries\sffamily #2 \par\nobreak}
  \lhead{}
  \chead{%
    \textit{\footnotesize XIV International Conference on Hadron Spectroscopy
      (\weblink[\textit{hadron2011}]{http://www.hadron2011.de}), 13-17 June 2011, Munich, Germany}%
  }
  \rhead{}
  \bigskip
  \begin{center}
    {#3} {#4}\ifthenelse{\equal{#6}{}}{}{\footnote{\weblink[#6]{mailto:#6}}}
    \ifthenelse{\equal{#7}{}}{}{#7} \\
    \textit{#5}
  \end{center}
  \bigskip
}

\renewcommand{\abstract}[1]{%
  \begin{center}
    \begin{minipage}{0.85\textwidth}
      \begin{footnotesize}
        #1
      \end{footnotesize}
    \end{minipage}
  \end{center}
  \bigskip
}

%% file: bonomi.tex
%
%
%
%
%
{  


%

\contribution[Hypernuclei Production by K$^-$ at rest]  
{Hypernuclei Production by K$^-$ at rest }  
{G.}{Bonomi}  
{Department of Mechanical and Industrial Engineering  \\
  via Branze 38, Brescia, ITALY \\
  \& INFN Pavia \\
  via Bassi 6, Pavia, ITALY}  
{germano.bonomi@ing.unibs.it}  
{}  
%

\abstract{%
The creation of a hypernucleus requires the injection of {\it strangeness} into the nucleus. This is possible in different ways, mainly using $\pi^+$ or K$^-$ beams on fixed targets. A  review of hypernuclei production by K$^-$ at rest is here presented. When a K$^-$ stops inside a nucleus it can undergo the so called "strangeness-exchange reaction", in which a neutron is replaced by a $\Lambda$ with the emission of a pion. By precisely studying the outgoing pions both the binding energy and the formation probability of the hypernuclei can be measured. New measurements from the FINUDA experiment on $^7$Li, $^9$Be, $^{13}$C and  $^{16}$O, coupled with previous measurements on $^{12}$C and $^{16}$O, allowed for the first time the study of the formation of hypernuclei as a function of the atomic mass number A. The new measurements also offered the possibility of disentangling the effects due to atomic wave-function of the captured K$^-$ from those due to the pion optical nuclear potential and from those due to the specific hypernuclear states. These new results on the study of the hypernuclei production by K$^-$ at rest are here presented and discussed.
}
%

\section{Introduction}

Hypernuclei are nuclear systems in which one or more nucleons are replaced by one (or more) hyperons \cite{uno}. The more known and studied for a long time (almost 60 years) are $\Lambda$-hypernuclei, in which one $\Lambda$-hyperon replaces a nucleon of the nucleus. In Fig. ~\ref{f:fig1}a) a simplified representation of the single-particle states of the nucleons for a $^{12}$C nucleus is shown. When the $\Lambda$ replaces a neutron [Fig.~\ref{f:fig1}b)] it can occupy different states [see Fig.~\ref{f:fig1} c) to f)]. When it has the same quantum numbers of the neutron it has replaced, the state is called "{\it substitutional}". 

\begin{figure}[htb]
  \begin{center}
    \includegraphics[width=\textwidth]{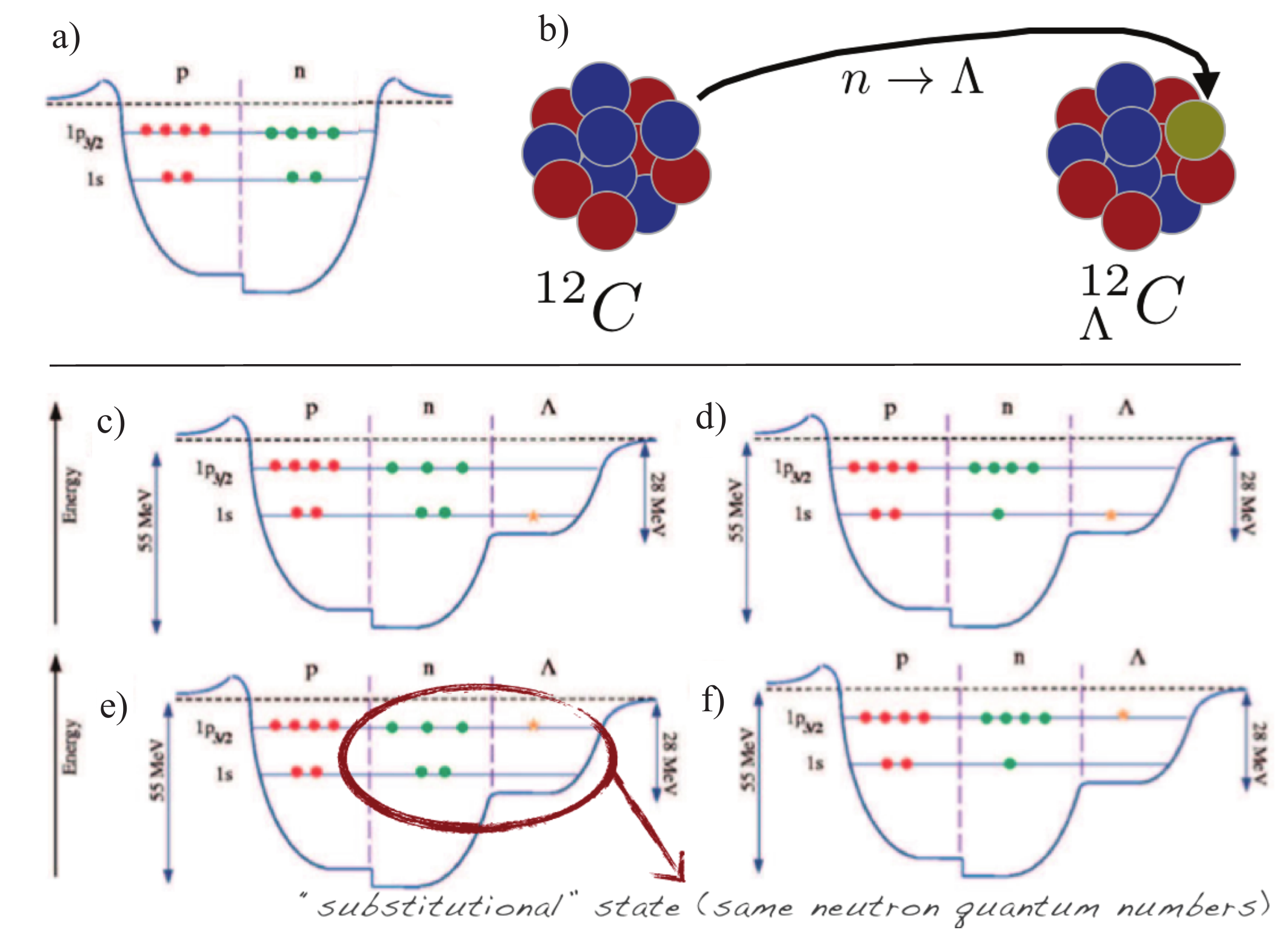}
    \caption{a) - Simplified representation of the single-particle states of the nucleons for a $^{12}$C nucleus. c) to f) - Same as a) when a $\Lambda$ replaced a neutron. State shown in e), when the $\Lambda$ has the same quantum numbers of the neutron it replaced, it is called {\it substitutional} state. Representations taken from \cite{bressani.report}.}
    \label{f:fig1}
  \end{center}
\end{figure}

Hypernuclei provide a unique laboratory suitable not only for studying nuclear structure in presence of a strange quark but also for probing weak interactions between baryons. Indeed hyperons-nucleons (YN) scattering experiments are difficult to perform and data are very limited. On the other hand from hypernuclear energy levels and theoretical models information about YN interaction can be extracted. Hypernuclei are also important to investigate dynamical changes of the nuclear structure induced by the added hyperon. In fact the $\Lambda$, reaching deep inside levels, it can attract the surrounding nucleons toward the interior ("{\it glue-like}" or "{\it core contraction}" effect), especially in light nuclei. Indeed the $\Lambda$-hyperon, since it does not suffer from Pauli blocking by other nucleons, it can penetrate into the nuclear interior and form deeply bound hypernuclear states. 

\section{Hypernuclei production}
There are different ways to bring the {\it strangeness} inside a nucleus. Up to now three different reactions have been used: 

  \begin{center}
    \begin{tabular}{l l l}
      1a)   &  ${\mathrm K}^-_{in flight/stop} + n \rightarrow \Lambda + \pi^- $     & \multirow{2}{*}{\it strangeness exchange reaction}    \\
      \vspace{0.3 cm}
      1b)   &  ${\mathrm K}^-_{in flight/stop} + p \rightarrow \Lambda + \pi^o$     &      \\
      \vspace{0.3 cm}
      2)   &  $\pi^+ + n \rightarrow \Lambda + {\mathrm K}^+$     &      {\it associated production} \\
      \vspace{0.3 cm}
      3)   &  $e + p \rightarrow e' + \Lambda + {\mathrm K}^+$     &      {\it electroproduction} \\

    \end{tabular}
    \label{t:reac}
  \end{center}

Each reaction has its own advantages and plays its role in a complete program of hypernuclear spectroscopy. Reaction 1a) at rest was the first to be used \cite{faessler}; reaction 2) then followed at BNL \cite{BNL1} and at KEK \cite{KEK1}, while reaction 3) is relatively new \cite{Jlab1}. The most important parameter in determining the differences of the distinct reactions is the momentum transfer. In Fig.~\ref{f:fig4} the relation between the beam momentum and the recoil/transfer momentum and between the momentum transfer and the hypernuclear cross section are shown. Low recoil momentum, like for the (K$^-_{inflight}$, $\pi^-$) reaction populates substitutional states, in which a nucleon is converted to a $\Lambda$ hyperon in the same orbit with no orbital angular momentum transfer. In this way it is difficult to populate the ground state. A large recoil momentum on the other hand can excite high-spin hypernuclear states with a nucleon-hole having large angular momentum and a $\Lambda$ hyperon having a small angular momentum, with the advantage to access more states. 

The pros of the (K$^-_{stop}$,$\pi^-$) reaction, if compared with the others, is the fact that it populates many states and it has a {\it high} formation probability compared to the associate production. On the other hand the kaon beam suffer energy struggling in the slowing down by {\it thick} absorbers and thus also the target must be relatively {\it thick} (some g/cm$^2$). This determines the principal drawbacks, like a large background and a limit on the energy resolution (the out-coming pions suffer multiple scattering inside the targets). 


A complete list of experiments on $\Lambda$ hypernuclear spectroscopy is shown in Fig.~\ref{f:fig3} \cite{hashimoto}. It can be seen that the K$^-_{stop}$ reaction was not so commonly used. Before the FINUDA experiment, which results will be presented in the following, the only other experiments that produced hypernuclei with the ${\mathrm K}^-_{stop} + n \rightarrow \Lambda + \pi^- $ reaction were two \cite{faessler,hayano}. Recently another experiment used the ${\mathrm K}^-_{stop} + p \rightarrow \Lambda + \pi^o$ reaction \cite{ahmed}. All these experiments used spectrometers designed for other purposes and modified to fit the needs of a hypernuclear apparatus. The FINUDA experiment on the other hand was specifically planned to produce and study hypernuclei. 

\begin{figure}[tb]
  \begin{center}
    \includegraphics[width=0.9\textwidth]{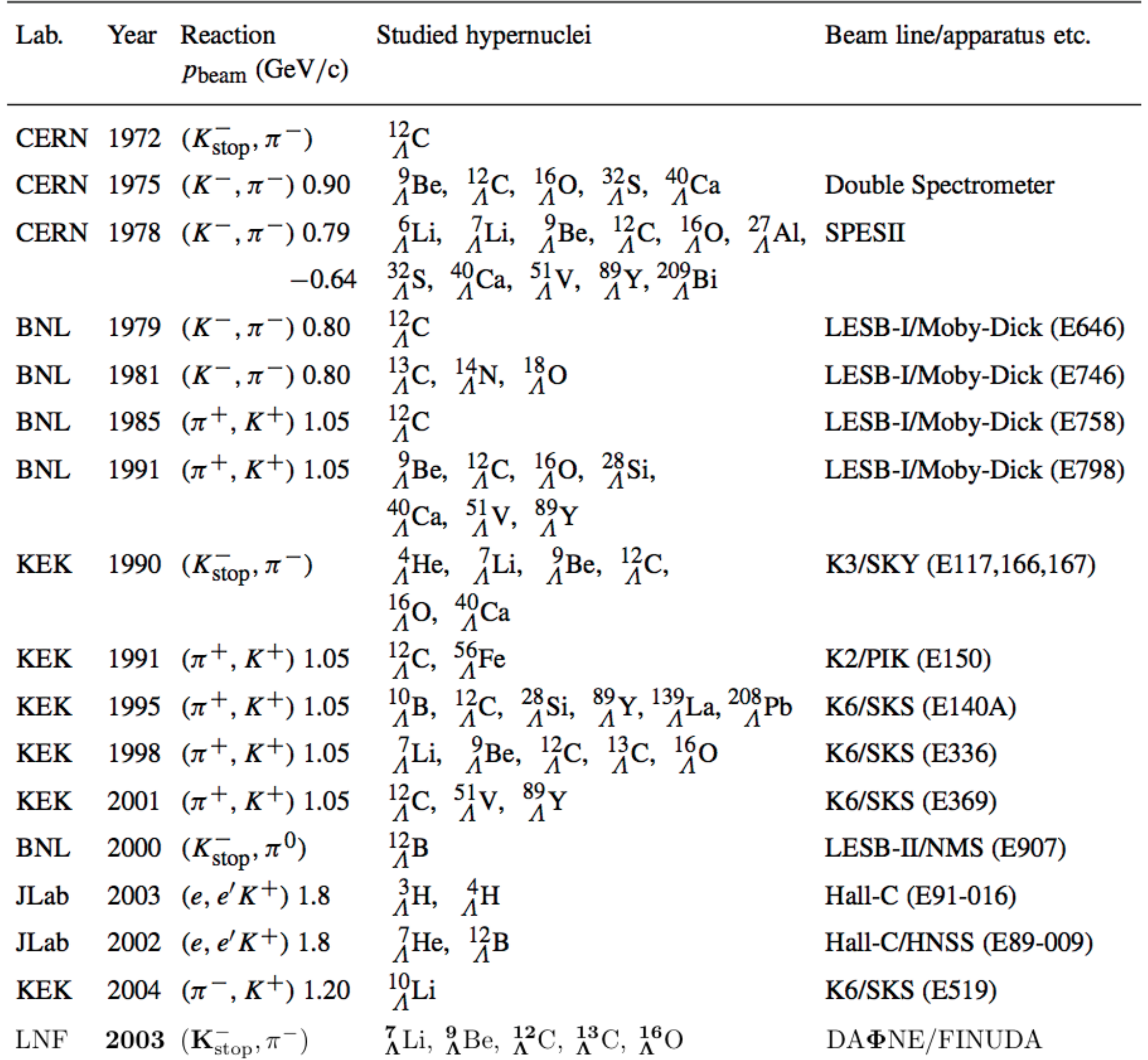}
    \caption{Experiments on $\Lambda$ hypernuclear spectroscopy (from \cite{hashimoto} with the addition of FINUDA).}
    \label{f:fig3}
  \end{center}
\end{figure}

\section{FINUDA hypernuclear spectroscopy}
FINUDA took data for few months in between 2003 and 2004 and 2006 and 2007 at the DA$\Phi$NE $e^+$-$e^-$ collider machine at the national laboratory of the Italian Institute of Nuclear Physics (INFN) in Frascati. The  $e^+$-$e^-$ collisions create the $\Phi$ meson at rest that decays, about 50 $\%$ of the times, into two kaons with  low kinetic energy ($\sim$ 16 MeV). The basic principle of FINDUA \cite{bressani} was to use such monochromatic source of low energy K$^-$'s for the production of hypernuclei. Since it is impossible to obtain such low energy beams in other ways (for example with fixed target experiments as done previously), FINUDA represented a real breakthrough for stopped kaons experiments. FINUDA in particular was characterized by important features, in particular it could: mount very thin targets (few tens of g/cm$^2$ compared to some g/cm$^2$ of previous experiments),  install up to 8 different targets in the same data taking (thus minimizing the systematics in comparing results from different nuclei), completely reconstruct an event with large acceptance (studying both the production and the decay of hypernuclei),  simultaneously track also the $\mu^+$ from the decay of the K$^+$ (which is generated in pair with the K$^-$) calibrating in this way the apparatus both for energy and rate measurements. Details about the FINUDA experimental setup can be found in \cite{C12, MWD, NMWD} and references therein.

\begin{figure}[tb]
  \begin{center}
    \includegraphics[width=0.9\textwidth]{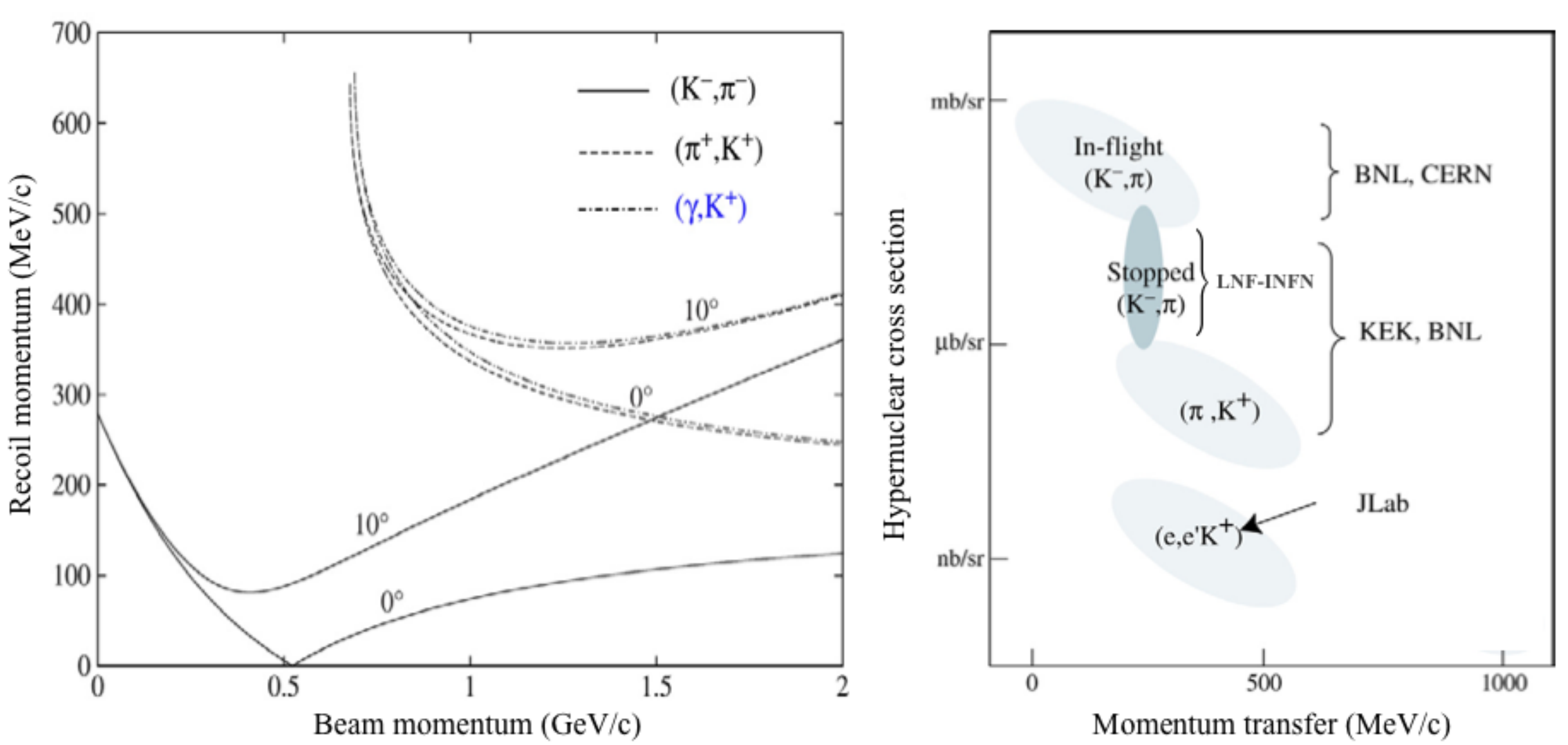}
    \caption{a) Recoil momentum as a function of the beam momentum (from \cite{hashimoto}). When the momentum transfer is 0 ({\it magic (beam) momentum}), the hypernucleus production is called recoilless. b) Hypernuclear cross section as a function of the momentum transfer (from \cite{bressani.report}). The production of hypernuclei by particles at rest is in reality defined by a capture rate (or formation probability) and not by cross section. Suitable normalization has been used.}
    \label{f:fig4}
  \end{center}
\end{figure}
The FINUDA apparatus was able to reconstruct charged particles coming out of the targets. Hypernuclear candidate events were selected by requiring the simultaneous presence of a K$^-$ stopped inside a target and of a $\pi^-$ originating from the same target. Details about the event selection and data analysis can be found elsewhere \cite{germano, C12}. Imposing the conservation of the energy and of the momentum, the tracking of the emerging $\pi^-$ permits to calculate the hypernucleus binding energy, defined as the difference between the mass of the hypernucleus and the sum of the masses of the core nucleus (original nucleus without a neutron) and of the neutron. For what concerns the background, some other reactions between the K$^-$ and the target can produce an emerging negative pion [$K^- (np) \rightarrow \Sigma^- p$  (followed by $\Sigma^- \rightarrow n \pi^-$ decay), $K^- n \rightarrow \Lambda \pi^- $ (so called {\it $\Lambda$-Quasi Free}), $K^- p \rightarrow \Sigma^- \pi^+$ (followed by $\Sigma^- \rightarrow n \pi^-$ decay)]. Another process that proved to be able to generate a $\pi^-$ momentum distribution that can overlap with the one of hypernuclear formation is the in-flight K$^-$ decay. All these reactions have been simulated with the FINUDA Monte Carlo  in order to account for the background of our hypernuclear signal. A sum of the distributions of the background reactions and of Gaussians, for the signal, was used to reproduced the overall data behaviour (see \cite{germano} for details). The output of the fit was the weight of the various contributions, the number of events and the mean of the Gaussians. The position of the peaks gives directly the binding energy value of the hypernuclear states created, with a total error of 0.3 MeV. A more important information can be extracted from the number of events in the peaks. Taking into account all the efficiencies involved in the detection and reconstruction, the formation probability per stopped K$^-$ can be calculated.

\section{Results and discussions}
The values of formation probabilities measured by the FINUDA experiments for $^7_{\Lambda}$Li,  $^9_{\Lambda}$Be and $^{13}_{\Lambda}$C and $^{16}_{\Lambda}$O are reported in a recent publication \cite{germano}. Only few measurements of formation probability have been performed previously. Following the first experiment on a $^{12}$C stopping target \cite{faessler}, measurement on some other nuclei ($^4$He, $^{12}$C and $^{16}$O) were subsequently performed \cite{hayano}. A low statistics measurement on the ($K^-_{stop}, \pi^o$) reaction on $^{12}$C was later published \cite{ahmed}. FINUDA also reported a first result on a $^{12}$C target \cite{C12}. Based on these measurements the formation probability was a decreasing function of the atomic mass number A, but some discrepancy appeared for example between the ground state formation probabilities measured by \cite{C12,hayano} and \cite{ahmed}. The new FINUDA results \cite{germano}, along with the one reported previously by FINUDA \cite{C12}, offer a complete set of measurements that can be compared one each other to extract the formation probability as a function of the atomic mass number A. Since they were measured in the same experiment and using the same experimental and reconstruction techniques, the effect of systematic errors is thus minimized.

\begin{figure}[htb]
  \begin{center}
    \includegraphics[width=\textwidth]{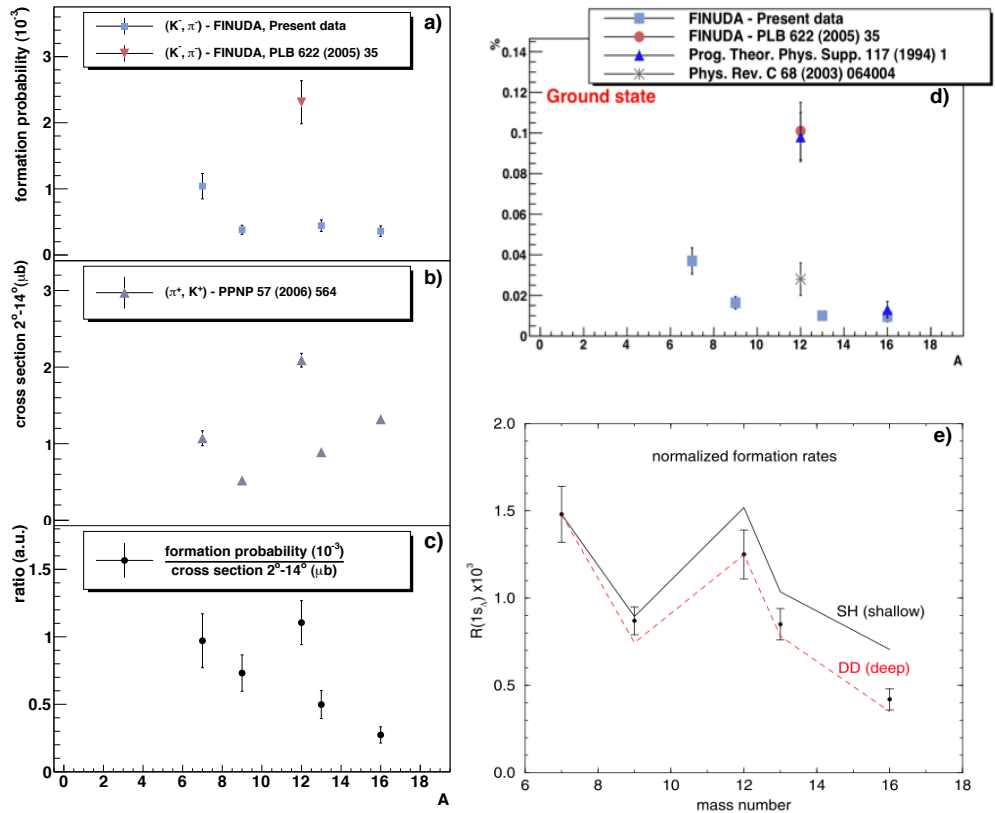}
    \caption{Formation probabilities from FINUDA \cite{germano} (a) and cross section from E336 \cite{hashimoto} (b) for bound states (see text for details) as a function of A. Ratio between the two c). Theoretical prediction by \cite{cieply} (e). Formation probabilities for the ground states (d).}
    \label{f:fig5}
  \end{center}
\end{figure}

Since each target has more than one hypernuclear state (the ground plus some excited), it is not easy to extract the A dependence from the results. The ground state formation probability, as shown in Fig.~\ref{f:fig5} d), can be used, but for some hypernuclei (namely $^7_{\Lambda}$Li and $^{12}_{\Lambda}$C) the first excited state is too close to be isolated. In order to consider only well defined hypernuclides, hypernuclear states below the threshold for the decay by proton emission have been selected. The results are summarized in  Fig.~\ref{f:fig5} a). A smoothly decreasing behaviour appears, with the exception of a strong enhancement corresponding to the formation of $^{12}_{\Lambda}$C. For comparison, in Fig.~\ref{f:fig5} b) the cross section measurement for the ($\pi^+$, K$^+$) production reaction \cite{hashimoto} is also shown. The ratio between such values and the (K$^-_{stop}$, $\pi^-$) formation probability ones (Fig.~\ref{f:fig5} c) changes by a factor of 5 from $^7$Li to $^{16}$O. Two conclusions can be then drawn. First of all, the two production reactions (K$^-_{stop}$, $\pi^-$)  and ($\pi^+$, K$^+$) show a distinct A dependence. Secondly, hypernuclei production on $^{12}$C deviates from the overall behaviour, being higher than all other neighbour nuclei. 

The new results \cite{germano} triggered a new study \cite{cieply} to reproduce experimental data to extract information on the K$^-$ nuclear potential V$_K$, important for kaon condensation in neutron-star matter, quasibound K-nuclear clusters, self-bound strange hadronic matter, etc. etc. The authors used the Distorted-Wave Impulse Approximation (DWIA) to calculate the formation probability for the target nuclei reported in \cite{germano}. Two different potentials were tested, namely a shallow (SH) and a deep (DD) ones with respectively $Re$ V$_K$ $\sim$ 50 MeV and $Re$ V$_K$ $\sim$ 180 MeV. The dependence on the nuclear density has also been taken into account. Although the calculated rates were about 15 $\%$ of the measured rates, the overall behaviour could be reproduced, the comparison slightly preferring the deep K$^-$ nuclear potential over the shallow one. 

\section{Conclusions}
The ${\mathrm K}^-_{stop} + n \rightarrow \Lambda + \pi^- $ strangeness exchange reaction was the first \cite{faessler} to be used for the creation of hypernuclei in {\it modern} (post emulsion/bubble chambers era). After having being used at BNL \cite{BNL1} and KEK \cite{KEK1}, it found its {\it best configuration} at the INFN-LNF with the FINUDA experiment. Hypernuclei formation probabilities for stopped kaons has been measured for p-shell nuclei and for the first time a study as a function of A has been performed \cite{germano}. The new results gave new inputs to the theory to extract useful information about the K$^-$ nuclear potential. 
No new experiment using the (K$^-_{stop}$, $\pi^-$) reaction is foreseen at the moment since future programs of hypernuclear physics, at JParc in Japan and at GSI in Germany, will be using different production methods. Complete reviews of hypernuclear physics can be found in \cite{bressani.report, hashimoto, chrien, bando}.



%

}  
